\documentclass[twocolumn,preprintnumbers,amsmath,amssymb,showpacs,superscriptaddress,prl,
]{revtex4} 

\usepackage{amsmath}
\usepackage{amssymb}
\usepackage{graphicx}

\begin{document}

\title{Generalized space and linear momentum operators in quantum {mechanics} }

\author{Bruno G. da Costa}
\email{bruno.costa@ifsertao-pe.edu.br} 
\affiliation{Instituto Federal de Educa\c c\~ao, Ci\^encia e Tecnologia do
             Sert\~ao Pernambucano, {\it Campus} Petrolina, 
             BR 407, km 08, 56314-520 Petrolina, Pernambuco, Brazil} 
\affiliation{Instituto de F\'isica, Universidade Federal da Bahia, Campus Universit\'ario de Ondina, 40170-115 Salvador,  Bahia, Brazil} 
\author{Ernesto P. Borges} 
\email{ernesto@ufba.br} 
\affiliation{Instituto de F\'isica, Universidade Federal da Bahia,
             Campus Universit\'ario de Ondina, 40170-115 Salvador,  Bahia, Brazil} 
\affiliation{National Institute of Science and Technology for Complex Systems, Brazil} 

\date{\today} 

\begin{abstract}
We propose a modification of a recently introduced generalized 
translation operator, by including a $q$-exponential factor,
which implies in the definition of a Hermitian deformed 
linear momentum operator $\hat{p}_q$,
and its canonically conjugate deformed position operator $\hat{x}_q$.
A canonical transformation leads the Hamiltonian of a position-dependent 
mass particle to another Hamiltonian of a particle with constant mass 
in a conservative force field of a deformed phase space. 
The equation of motion for the classical phase space 
may be expressed in terms of the generalized dual $q$-derivative.
A position-dependent mass confined in an infinite square potential well
is shown as an instance.  
Uncertainty and correspondence principles are analyzed.
\end{abstract} 



\maketitle

Systems consisting of particles with position-dependent mass (PDM)
have been discussed by several researchers since few past decades.
Applications of such systems may be found in
semiconductor theory \cite{vonross_1983},
${}^4$He impurity in homogeneous liquid ${}^3$He \cite{Saavedra_1994},
nonlinear optics \cite{Khordad_2012}, studies of inversion potential for
NH$_{3}$ in density functional theory (DFT) \cite{Aquino_1998},  
particle physics  \cite{Bethe_1986},
and astrophysics \cite{Richstone_1982}. 

Recently, Costa Filho {\it et al.} \cite{costa-filho-2011,costa-filho-2013}
have introduced a generalized translation operator which produces infinitesimal 
displacements related to the $q$-algebra \cite{lemans,borges_2004}, 
{\it i.e.},
\begin{equation}
\label{eq:translate}
  \hat{T}_{\gamma} (\varepsilon )|x\rangle 
  \equiv |{x + \varepsilon + \gamma x\varepsilon}\rangle , 
\end{equation}
where $\gamma$ is a parameter with dimension of inverse length. 
This operator leads to a generator operator of spatial translations  
corresponding to a position-dependent linear momentum
given by $ \hat{p}_\gamma = (\hat{1}+\gamma \hat{x}) \hat{p} $,
and consequently a particle with position-dependent mass.
This operator was used to solve problems of particles with 
position-dependent mass in the quantum formalism. 
This deformed momentum operator is not Hermitian, 
which led Mazharimousavi \cite{mazharimousavi} 
to introduce a modification in its definition.
Other generalizations have been also appeared in the literature, particularly
a nonlinear version of Schr\"odinger, Klein-Gordon, and Dirac equations 
\cite{Nobre-RegoMonteiro-Tsallis-2011,Nobre-RegoMonteiro-Tsallis-2012,
Plastino-Tsallis}.

We introduce 
a nonnormalized generalized phase factor in Eq.\ (\ref{eq:translate}) as
\begin{eqnarray}
\label{eq:translate-modif}
   \hat{T}_{q} (\varepsilon )|x\rangle &\equiv&
   \exp_{q}\left[ \frac{ig(x)\varepsilon}{\hbar}\right] 
   \left |{x + \varepsilon + \frac{1-q}{\xi} x\varepsilon}\right\rangle
\nonumber \\
&=& 
   \exp_{q}\left[ \frac{ig(x)\varepsilon}{\hbar}\right] 
   \left |{\xi (\tilde{x} \oplus_q \tilde{\varepsilon}}\right) \rangle,
\end{eqnarray}
where $g(x)$ is a continuous function with dimension of linear momentum
($g(x)=0$ recovers Eq.(\ref{eq:translate})),
$\varepsilon$ is an infinitesimal displacement,
the symbol $\oplus_q$ represents the $q$-addition operator,
$a \oplus_q b = a + b + (1-q)ab$
\cite{lemans,borges_2004},
$\xi$ is a characteristic length,
$\tilde{x}\equiv x/\xi$ is the dimensionless position,
and the dimensionless parameter $q$ controls the generalization of the
exponential function
$
\exp_q{x} \equiv [1 + (1-q)x]_+^{1/(1-q)}, 
$
with $[A]_+ \equiv \max\{A,0\}$ \cite{ct-quimicanova}.
The symbol $\gamma$ in Eq.~(\ref{eq:translate})
(as it appears in \cite{costa-filho-2011})
has been here changed to $\gamma_q\equiv(1-q)/\xi$, 
once the $q$-addition shall be used with dimensionless variables.

The $q$-exponential of an imaginary number yields generalized trigonometric
functions \cite{borges_trig_1998}, and it can be written as
$
 \exp_q(\pm ix) = \rho_q(x) \exp_1(\pm ix),
$
with $\rho_q^2(x)=\exp_q(ix)\exp_q(-ix)= \exp_q[(1-q)x^2]$
($x \in \mathbb{R}$);
$\rho_q(x)$ 
is the norm of the $q$-exponential.
$q=1$ recovers the usual exponential function,
and the $q$-exponential factor reduces to a usual phase factor
with unitary norm.
The $q$-exponential function satisfies
\begin{equation}
 \label{eq:exp_q(a)exp_q(b)}
  \exp_q{(a)}\exp_q{(b)} = \exp_q(a \oplus_q b)
\end{equation}
where $a$ and $b$ are two dimensionless quantities.

Similarly to the operator defined by Eq.~(\ref{eq:translate}), 
$\hat{T}_{q}(\varepsilon )$ also forms a group, {\it i.e.},
\begin{equation}
\label{eq:translate-property1}
   \hat{T}_{q}(\xi d\tilde{x}_1)\hat{T}_{q}(\xi d\tilde{x}_2) |0\rangle = 
    \hat{T}_{q}(\xi (d\tilde{x}_1 \oplus_{q} d\tilde{x}_2)) |0\rangle.
\end{equation}
%
Application of the operator $\hat{T}_q(\varepsilon)$
on state $|0\rangle$, repeated $n$ times, leads to
\begin{equation}
\label{eq:translate-property2}
   \hat{T}_{q}^n(\varepsilon)|0\rangle = 
   \exp_{q}\left[ n\odot _q \frac{ig(x)\varepsilon}{\hbar}\right] 
   |n\odot_{q}\varepsilon \rangle,
\end{equation}
where $n\odot_{q} x$ is a generalized product \cite{borges_2004}:
\begin{equation}
\label{eq:n-repeated-addition}
   n\odot_{q} x = 
   \frac{1}{1-q} \Bigl\{ \bigl[ 1+(1-q)x \bigr] ^n - 1 \Bigr\}.
\end{equation}
(Not to confound the generalized product $n\odot_{q} x$
with another generalization, frequently known as $q$-product,
$a \otimes_q b \equiv [a^{1-q}+b^{1-q}-1]^{1/(1-q)}$
\cite{lemans,borges_2004}.)
This expression may be analytically extended for $n\in \mathbb{R}$.
Particularly, for $x = 1$, $n\odot_{q} 1$ is identified with the
Heine deformed number
(see \cite{borges_q_numbers_2009}, and also \cite{quantumgroups-1994}
for a possible connection with quantum groups).

Let $|\psi_{\varepsilon} \rangle \equiv \hat{T}_q(\varepsilon)|\psi\rangle$.
The effect of the operator 
$\hat{T}_{q}(\varepsilon )$ on state $|\psi\rangle$ is
\begin{eqnarray}
|\psi_{\varepsilon} \rangle
 &=&  
\hat{T}_{q}(\varepsilon)\int |x\rangle \langle x|\psi \rangle dx 
\nonumber\\
&=&  
\int \exp_{q} \left[ \frac{ig(x)\varepsilon}{\hbar}\right]
|\xi (\tilde{x}\oplus_{q}\tilde{\varepsilon}) \rangle \langle x |\psi \rangle dx.
\end{eqnarray}
If 
$\psi_{\varepsilon}(x) \equiv 
                       \langle x|\hat{T}_{q}(\varepsilon ) |\psi\rangle$,
we have
\begin{equation}
  \label{eq:function-modified}
  \psi_{\varepsilon}(x) = 
  \exp_{q} \left[
 \frac{i\varepsilon}{\hbar}g\left(
    \xi (\tilde{x}\ominus_q \tilde{\varepsilon})
\right)
\right]
  \frac{\psi(\xi (\tilde{x}\ominus_q \tilde{\varepsilon}))}
       {1+\gamma_q\varepsilon}.
\end{equation}
where $\psi(x) = \langle x|\psi \rangle $, 
and the $q$-difference operator is defined as
$a\ominus_q b = \frac{a-b}{1+(1-q)b}$, with $b\ne 1/(q-1)$
(see \cite{lemans,borges_2004} for details).

We forward a simple application: 
let $|\psi\rangle $ be a normalized Gaussian state 
with width $\sigma$, and centered at $x=0$,
in the coordinate basis $ \{|x\rangle\} $. 
The effect of $\hat{T}_q(\varepsilon)$ on $|\psi\rangle$ leads to
\begin{equation}
  \psi_\varepsilon(x) =
    \frac{ e^{-(x-\varepsilon )^2/2\sigma_q^2}}{\sigma_q\sqrt{2\pi }}
    \exp_{q}
    \left[
    \frac{i\varepsilon}{\hbar}g\left(\frac{x-\varepsilon}{1+\gamma_q\varepsilon}\right)
                      \right],
\end{equation}
with $\sigma_q = \sigma(1+\gamma_q\varepsilon)$,
which is normalized up to $\mathcal{O}(\varepsilon)$.
If $g(x)=0$, the effect of $\hat{T}_q(\varepsilon)$ on a Gaussian
packet yields a shift in $x$ 
and an increase in its width for $1 + \gamma_q\varepsilon > 0$, 
or a decrease for $1 + \gamma_q\varepsilon < 0$.
Let the expected value 
$  {\langle \hat{x} \rangle}_{\varepsilon}  = \int
  \psi^{\ast}_\varepsilon (x') x' \psi_\varepsilon (x') dx'  $,
by changing the variable of integration 
$x = \xi (\tilde{x}' \ominus_q \tilde{\varepsilon})$,  we have
\begin{equation}
  {\langle \hat{x} \rangle}_{\varepsilon} =
         \int dx
      \psi^{\ast}(x)(x + \varepsilon + \gamma_q x \varepsilon )
      \psi (x)\frac{e_{q}^{(1-q){\varepsilon}^2 g^2(x)/{\hbar}^2}}{1+\gamma_q \varepsilon }.
\end{equation}
The first order approximation in $\varepsilon$ is represented by 
the $q$-addition:
\begin{equation}  
   {\langle \hat{x} \rangle}_{\varepsilon} =   
   {\langle \hat{x} \rangle} + \varepsilon 
   + \gamma_q{\langle \hat{x} \rangle}\varepsilon. 
\end{equation}  

Equation (\ref{eq:exp_q(a)exp_q(b)}) naturally suggests the definition
\begin{equation}
\label{eq:gerator-of-translation}
   \hat{T}_q ( \varepsilon ) \equiv  
    \exp_q \left( -\frac{i\varepsilon \hat p_{q}}{\hbar}\right),
\end{equation}
$\hat{p}_q $ is the generator of generalized infinitesimal translations.
Expanding $\hat{T}_q(\varepsilon)$,
and $\psi_\varepsilon(x)$ (Eq.~(\ref{eq:function-modified})),
up to the first order in $\varepsilon$, we get
\begin{equation}
\label{eq:expansion-translate}
   \hat{T}_q(\varepsilon) =  
    \hat{1}-\frac{i\varepsilon \hat{p}_{q}}{\hbar} + ...,
\end{equation}
\begin{eqnarray}
 \label{eq:expansion-psi-displacement}
 \psi_\varepsilon(x) &=&  (1-\gamma_q \varepsilon+ ...)(1 + \varepsilon A + ....) 
 \nonumber \\           
 && \times 
    \left[ \psi(x) - \varepsilon (1 + \gamma_q x)\frac{d\psi}{dx} +... \right],
\end{eqnarray}
where $A$ is a constant taken from the expansion of 
$\exp_q (ig(x)\varepsilon / \hbar)$ in powers of $\varepsilon$,
and we have
\begin{equation}
\label{eq:operator-momentum-wiht-A}
	\langle x|\hat{p}_q |\psi \rangle = 
    -i\hbar \frac{d}{dx} [(1+\gamma_q x)\psi(x) ]+ i \hbar A \psi(x).
\end{equation}
Imposition that  $\hat{p}_q$  is Hermitian implies $A=\gamma_q/2$, then
\begin{equation}
\label{eq:operator-momentum-with-A=gamma/2}
	  \hat{p}_q = 
      \hat{p}(\hat{1}+\gamma_q \hat{x})+\frac{1}{2}i\hbar \gamma_q\hat{1} =
      (\hat{1}+\gamma_q \hat{x})\hat{p} - \frac{1}{2} i\hbar \gamma_q \hat{1},
\end{equation}
{\it i.e.,}
\begin{equation}
 \label{eq:operator-momentum-generalized}
 \hat{p}_q = \frac{(\hat{1}+\gamma_q \hat{x})\hat{p}}{2} + 
                  \frac{\hat{p}(\hat{1}+\gamma_q \hat{x})}{2},
\end{equation}
with
$ [\hat{x}, \hat{p}] = i\hbar \hat{1} $.

We introduce a generalized space operator $\hat{x}_q $ such that
$[\hat{x}_q, \hat{p}_q] = i\hbar \hat{1}$.
Recalling the property
$ [f(\hat{x}) , \hat{p}] = i\hbar f'(\hat{x})$,
with $ \hat{x}_q = f(\hat{x}) $,
we arrive at
\begin{equation}
 \label{eq:operator-position-generalized}
 \hat{x}_q = \frac{\ln (\hat{1}+\gamma_q \hat{x})}{\gamma_q} 
                = \xi \ln[\exp_q ( \hat{x}/\xi) ].
\end{equation}
The transformation (\ref{eq:operator-position-generalized}) 
has already appeared in a different context,
as the real part of a transformation of a complex number $z$ 
into a kind of generalized complex number $\zeta_q=\ln \exp_q z$, 
and this allows the $q$-Euler formula to be expressed as 
$\exp_q z = \exp_1 \zeta_q$ \cite{borges_trig_1998}.
Even before that, the transformation (\ref{eq:operator-position-generalized}) 
had also appeared connecting Tsallis (nonadditive) entropy 
with R\'enyi (additive) entropy \cite{Entropia-Tsallis}.

According to Ehrenfest's theorem, the time evolution of the expectation values
of the space $\hat{x}$ and linear momentum $\hat{p}$ operators are given 
respectively by
\begin{subequations}
\begin{equation}
 \label{eq:d<x>/dt}
 \frac{d \langle \hat{x} \rangle}{dt} = 
  \frac{\langle (\hat{1}+\gamma_q \hat{x})^2 \hat{p} \rangle}{2m} 
  + \frac{\langle \hat{p} (\hat{1}+\gamma_q \hat{x})^2 \rangle}{2m},
\end{equation}
and
\begin{equation}
 \label{eq:d<p>/dt}
 \frac{d \langle \hat{p} \rangle}{dt} = 
  - \Biggl\langle \frac{dV}{d\hat{x}} \Biggl\rangle
  - \frac{\gamma_q \langle (\hat{1}+\gamma_q \hat{x}) \hat{p}^2 \rangle}{2m} 
  - \frac{\gamma_q \langle \hat{p}^2 (\hat{1}+\gamma_q \hat{x}) \rangle}{2m},
\end{equation}
\end{subequations}
where we have used the following commutation relations:
\begin{equation}
  [\hat{x},\hat{p}^2_q] = i \hbar (\hat{1}+\gamma_q \hat{x})^2 \hat{p}
                        + i \hbar \hat{p} (\hat{1}+\gamma_q \hat{x})^2,
\end{equation}
and
\begin{equation}
  [\hat{p},\hat{p}^2_q] = -i \hbar\gamma_q (\hat{1}+\gamma_q \hat{x}) \hat{p}^2
                          -i \hbar\gamma_q \hat{p}^2 (\hat{1}+\gamma_q \hat{x}).
\end{equation}

The operators $ \hat x_q $ and $ \hat p_q $ present the following 
classical analogs:
\begin{subequations}
 \label{eq:classic}
 \begin{equation}
 \label{eq:momentum_generalized_classic}
 p_q = (1 + \gamma_q x)p,
 \end{equation}
and
 \begin{equation}
 \label{eq:space_generalized_classic}
  x_q = \frac{\ln (1+\gamma_q x)}{\gamma_q} = 
 \xi \ln \left[ \exp_{q} (x / \xi) \right],
 \end{equation}
\end{subequations}
with $ \{x_q, p_q\}_{(x, p)} = 1 $. 
The generating function of the canonical transformations given by
Eq.'s (\ref{eq:classic}) is
$ \Phi(x_q, p) = - p(e^{\gamma_q x_q} - 1)/ \gamma_q $.

As an application, let us address a constant mass particle 
and linear momentum $ p_q $ under the influence
of a conservative force with potential $V(x_q)$,
whose Hamiltonian is 
\begin{equation}
\label{eq:hamiltonian_xgamma_pgamma}
 K(x_q , p_q ) = \frac{p_q ^2}{2m} +  V(x_q ).
\end{equation}

The canonical transformations (\ref{eq:classic}) lead to the new Hamiltonian
(see, for instance, \cite{cruz_y_cruz_2013})
\begin{equation}
\label{eq:hamiltonian_x_p}
 H(x , p) = \frac{p^2}{2m(x)} +  V(x),
\end{equation}
where the particle mass depends on the position $x$ as
\begin{equation}
\label{eq:m(x)}
 m(x) = \frac{m}{(1+\gamma_q x)^2}.
\end{equation}
The equation of motion is 
\begin{equation}
 \label{eq:dp/dt}
 \dot{p} = - \frac{\gamma_q (1+\gamma_q x)p^2}{m} - \frac{dV(x)}{dx},
\end{equation}
with
$p = m(x) \dot{x}$,
thus
\begin{eqnarray} \label{eq:equation-of-motion} 
m \left[ \frac{\ddot{x}}{(1+\gamma_q x)^2} - 
    \frac{\gamma_q \dot{x}^2}{(1+\gamma_q x)^3}  \right]  = - \frac{dV(x)}{dx}.
\end{eqnarray}
This equation may be conveniently rewritten as
\begin{equation}
\label{eq:second_newton_law_generalized}
       m \widetilde{D}^2_{\gamma_q,t}   x (t) = F(x),
\end{equation}
{\it i.e.}, a deformed Newton's law for a space with nonlinear displacements,
where $\widetilde{D}_{q,u} f(u)$ is the dual $q$-derivative, 
defined as 
$\widetilde{D}_{q,u} f(u) 
\equiv \lim_{u'\to u}\frac{f(u')\ominus_q f(u)}{u'- u} 
= \frac{1}{1+(1-q)f(u)} \frac{df(u)}{du}$
\cite{borges_2004}.
The second $q$-derivative must be taken as
\begin{equation}
  \widetilde{D}_{q,u}^2 f(u) 
        = \frac{1}{1+(1-q)f(u)} 
          \frac{d}{du}
             \left[
                   \frac{1}{1+(1-q)f(u)} \frac{df}{du}
             \right],
\end{equation}
similarly to what was done in the (different) generalized
derivative introduced by \cite{Nobre-RegoMonteiro-Tsallis-2011}.

The generalized displacement of a
position-dependent mass in a usual space ($d_q x$) is mapped into 
a constant mass in a deformed space 
with usual displacement ($d x_q$):
$d_q x \equiv
    {\xi} \left[ \left( \frac{x+dx}{\xi} \right) 
    \ominus_q 
    \left( \frac{x}{\xi} \right) \right] 
    = \frac{dx}{1+\gamma_q x} \equiv d x_q$. 
The temporal evolution is governed by the generalized dual derivative,
$\widetilde{D}_{\gamma_q,t} x = \frac{1}{1+\gamma_q x}\frac{dx}{dt}$.

The probability 
$ P_{\mbox{\footnotesize classic}}dx \propto {dx}/{v} $
to find a classical particle with 
position-dependent mass given by Eq.\ (\ref{eq:m(x)}),
between $x$ and $x + dx$, constrained to $0 \leq x \leq L$, 
and free of forces, is
\begin{equation}
 \label{eq:probability_classic}
 P_{\mbox{\footnotesize classic}}dx 
             = \frac{\gamma_q}{(1+\gamma_q x)\ln (1 + \gamma_q L)}dx.
\end{equation}
Note that the probability density $P_{\mbox{\footnotesize classic}}$ 
is independent of the initial condition, and the uniform distribution 
$P_{\mbox{\footnotesize classic}}\rightarrow 1/L $ is recovered 
as $\gamma_q \to 0$.

The first and second moments of the classical distribution according to
position and momentum are
\begin{subequations}
\label{eq:classic_first_second_moment}
 \begin{equation}
  \label{eq:xmed_classic}
  \overline{x} 
   = \frac{ \gamma_q L - \ln (1+\gamma_q L)}{\gamma_q \ln (1+\gamma_q L)},
 \end{equation}
 \begin{eqnarray} 
 \label{eq:xquadmed_classic} 
  \overline{x^2} 
                 & =  & \frac{\gamma_q ^2 L^2 - 2 \gamma_q L + 2\ln (1+\gamma_q L)}
                             {2\gamma_q ^2 \ln (1+\gamma_q L)},
 \end{eqnarray}
 \begin{equation}
 \label{eq:pmed_classic}
  \overline{p} = 0,
 \end{equation}
and
 \begin{equation}
 \label{eq:pquadmed_classic}
  \overline{p^2} = 
 2mE \frac{[(1 + \gamma_q L)^2 - 1]}{2(1+\gamma_q L)^2 \ln (1+\gamma_q L)},
 \end{equation}
\end{subequations}
where 
$\lim_{\gamma_q \to 0} \overline{x} = L/2 $,
$\lim_{\gamma_q \to 0}\overline{x^2} = L^2/3$,
$\lim_{\gamma_q \to 0}\overline{p^2} = 2mE$,
and $E$ is the energy of the particle.

Consider a system described by the Hamiltonian operator $\hat{K}$ 
at coordinate basis $\{|\hat{x}_q \rangle\} $. 
The time independent Schr\"odinger equation 
for a particle in a null potential at the basis $\{|\hat{x}_q \rangle\} $ is
\begin{equation}
 \label{eq:equation-of-schroedinger-potencial-null}
   \frac{1}{2m}\hat{p}_q^2 |\psi \rangle = E|\psi \rangle.
\end{equation}
Using Eq.~(\ref{eq:operator-momentum-generalized}), we have:
\begin{eqnarray}
 \label{eq:equation-diferential-schr}
   -\frac{{(1+\gamma_q x)}^2{\hbar}^2}{2m} \frac{d^2\psi}{dx^2}
   -\frac{{\hbar}^2\gamma_q( 1+\gamma_q x)}{m}\frac{d\psi}{dx} &
\nonumber  \\ 
-\frac{{\hbar}^2\gamma_q^2}{8m}\psi(x) = E\psi(x) , &
\end{eqnarray}
which can be rewritten in the form
\begin{equation}
 \label{eq:equation-diferential-mathe}
   u^2\frac{d^2\psi(u)}{du^2}+au\frac{d\psi(u)}{du} + b\psi(u) = 0,
\end{equation}
with $ u(x) = 1 + \gamma_q x $, $a = 2$, and
$ {b = \frac{2m}{\hbar^2}\left( E + \frac{\hbar^2\gamma_q^2}{8m} \right)}$. 

Similarly to what was done in
\cite{costa-filho-2011} and \cite{mazharimousavi}, 
Eq.~(\ref{eq:equation-diferential-schr}) corresponds 
to a position-dependent mass particle according to Eq.\ (\ref{eq:m(x)}).
The solution of Eq.~(\ref{eq:equation-diferential-schr}) is given by
\begin{eqnarray}
 \label{eq:autofunction-generalized}
   \psi(x) &=& \frac{\psi_0}{\sqrt{1+ \gamma_q x}}
    \exp \left[\pm \frac{ik}{\gamma_q} \ln (1+\gamma_q x) \right] 
\nonumber \\
           &=& \frac{\psi_0}{\sqrt{1+ (1-q) x/\xi}}
   \left[\exp_q(x/\xi)\right]^{\pm ik\xi}
\end{eqnarray}
and presents a singularity at $x = -1/\gamma_q$. 

For of a particle inside an infinite square potential well between $x = 0$ and $x = L$, 
the eigenfunctions and energies of the particle are respectively given by
\begin{equation}
 \label{eq:psi_n}
   \psi_n(x) = \frac{A_{q,n}}{\sqrt{1+ \gamma_q x}}
               \sin \left[ \frac{k_{q,n}}{\gamma_q}\ln (1+ \gamma_q x) \right ] 
\end{equation}
for $0 \leq x \leq L$, and $\psi_n(x) = 0$ otherwise,
and
\begin{equation}
 \label{eq:energy}
    E_n = \frac{\hbar^2 \pi^2 \gamma_q^2 n^2}{2m \ln^2 (1+\gamma_q L)},
\end{equation}
with $A_{q,n}^2 = 2\gamma_q /\ln(1 + \gamma_q L)$, 
$ k_{q,n} =  n \pi \gamma_q / \ln(1+\gamma_q L) $ ($n$ is a integer number).
The wave function differs from those found in 
\cite{costa-filho-2011}, and \cite{mazharimousavi}, 
though it is similar to that one obtained in \cite{schmidt-2006}.
Nevertheless, the energy levels are the same of those in \cite{costa-filho-2011}. 

\begin{figure}[ht]
\centering
\begin{minipage}[h]{0.48\linewidth}
\includegraphics[width=\linewidth]{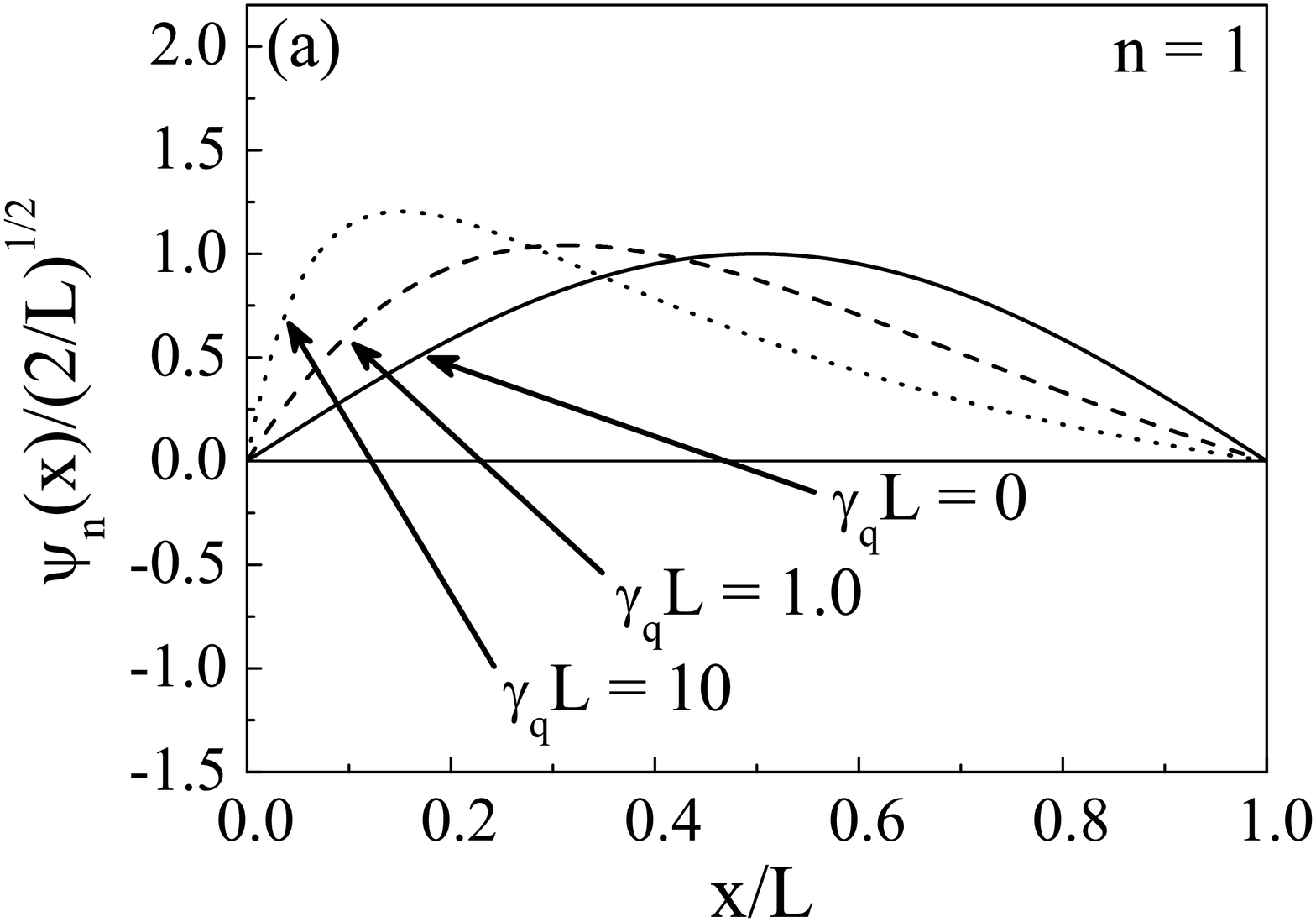}
\end{minipage} 
\begin{minipage}[h]{0.48\linewidth}
\includegraphics[width=\linewidth]{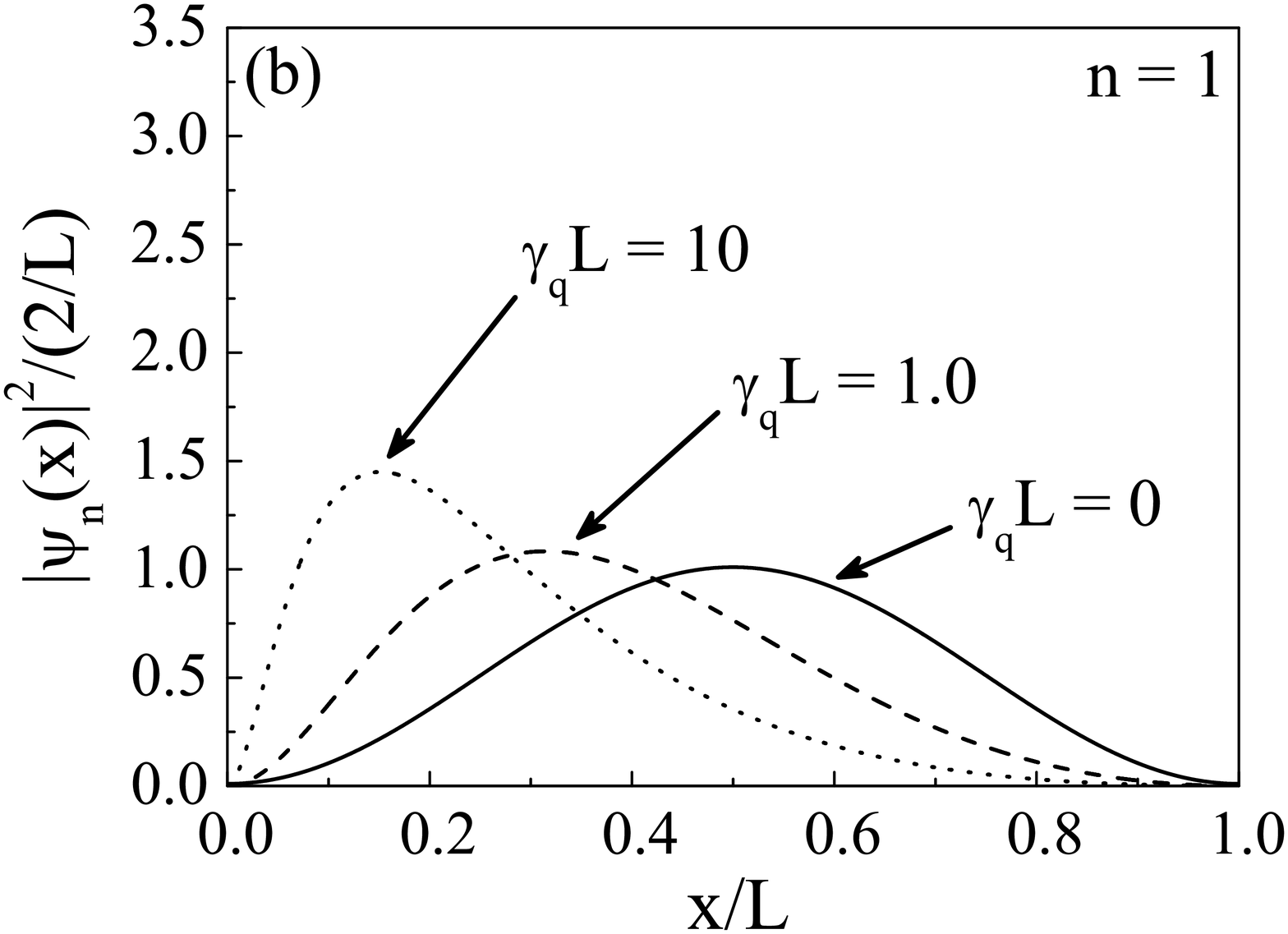}
\end{minipage} \\
\begin{minipage}[h]{0.48\linewidth}
\includegraphics[width=\linewidth]{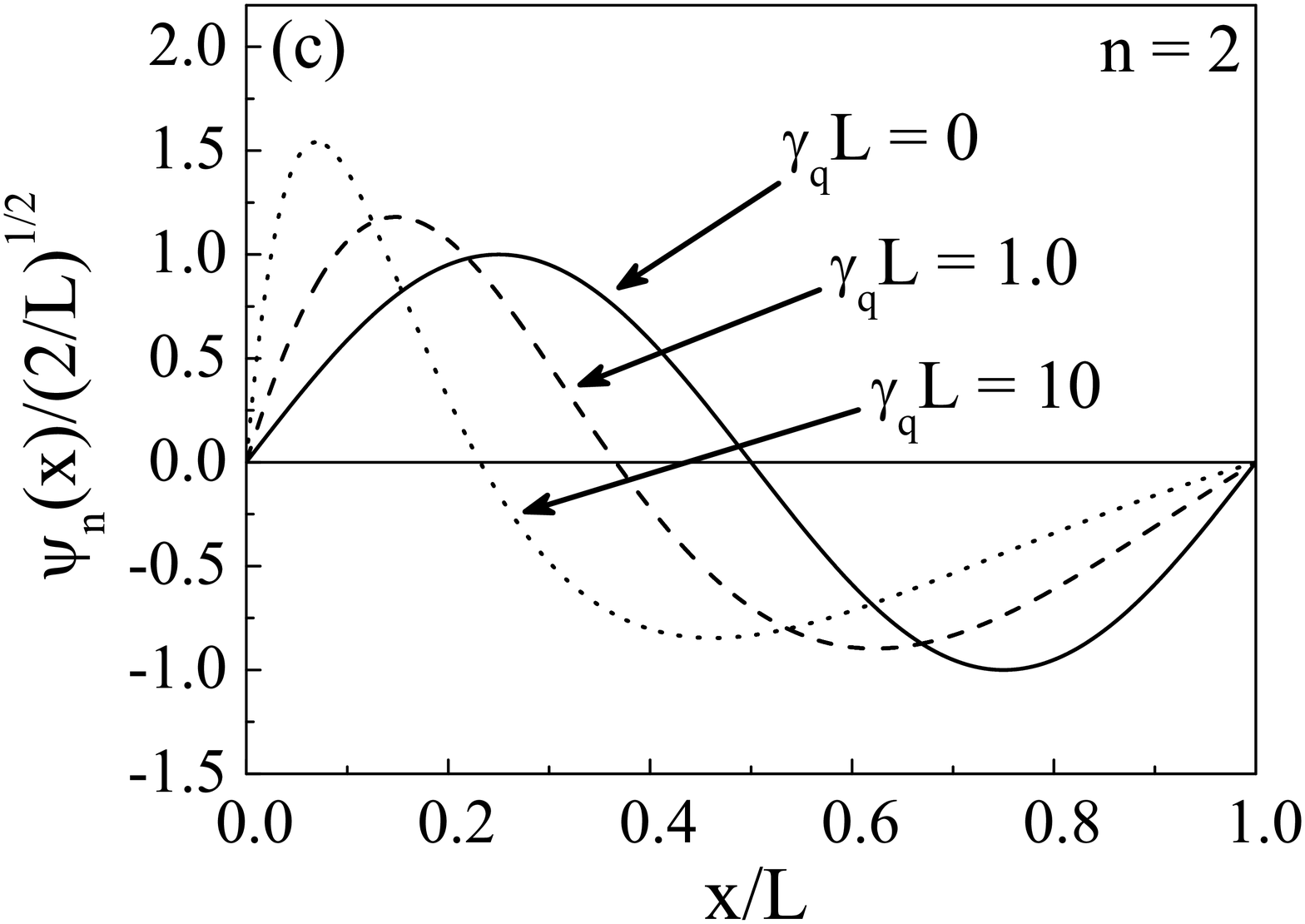}
\end{minipage} 
\begin{minipage}[h]{0.48\linewidth}
\includegraphics[width=\linewidth]{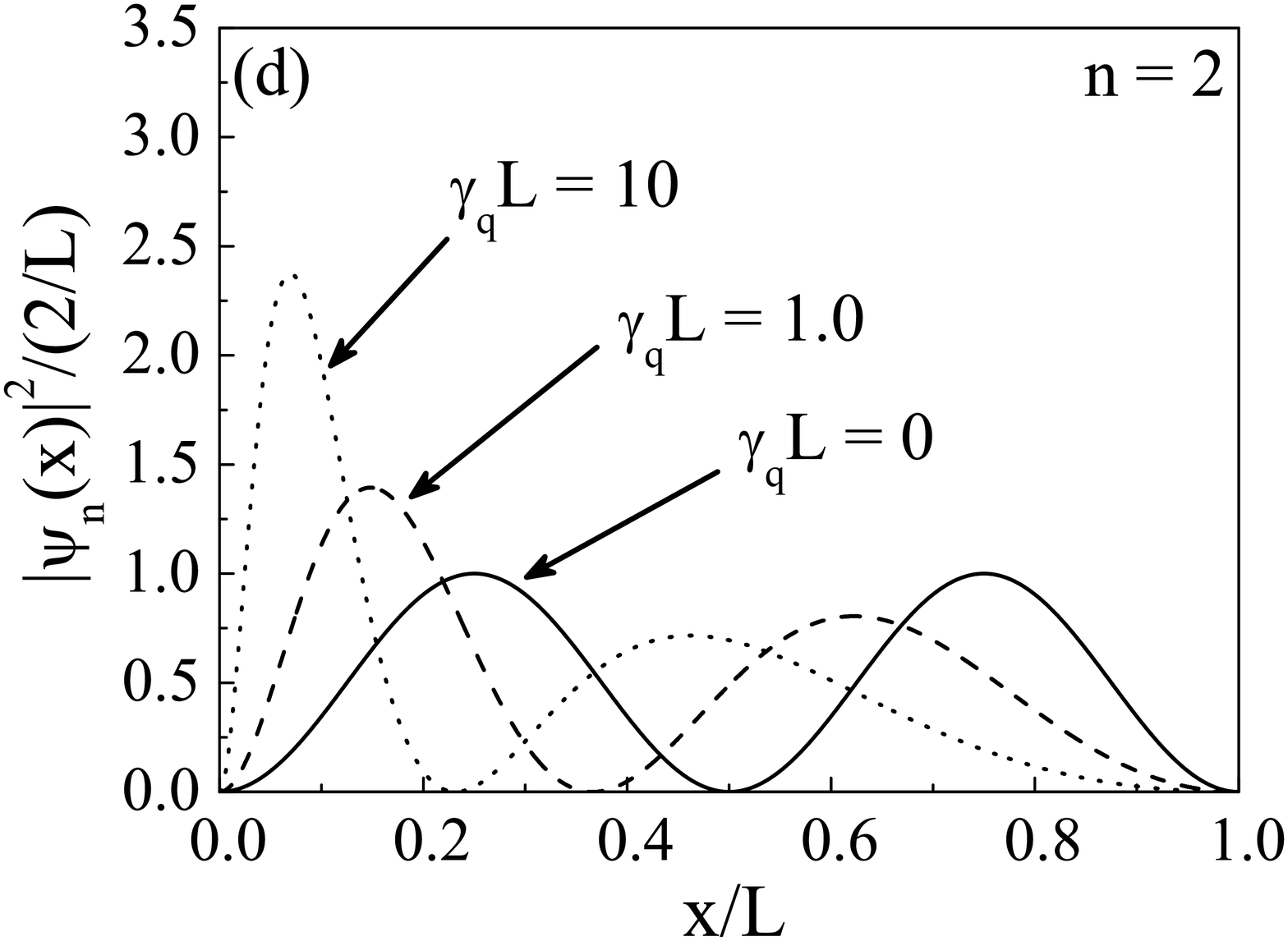}
\end{minipage}\\
\begin{minipage}[h]{0.48\linewidth}
\includegraphics[width=\linewidth]{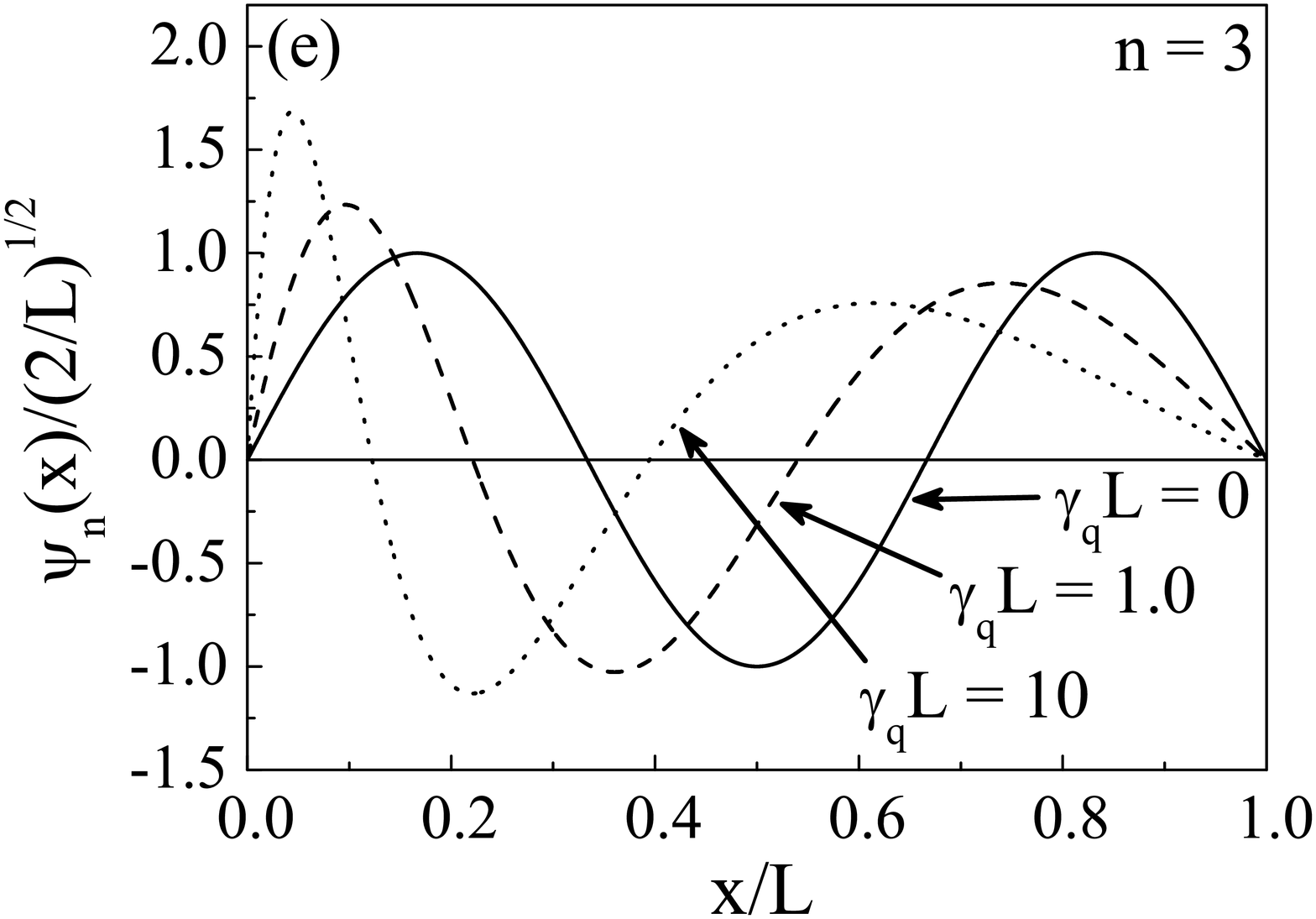}
\end{minipage}
\begin{minipage}[h]{0.48\linewidth}
\includegraphics[width=\linewidth]{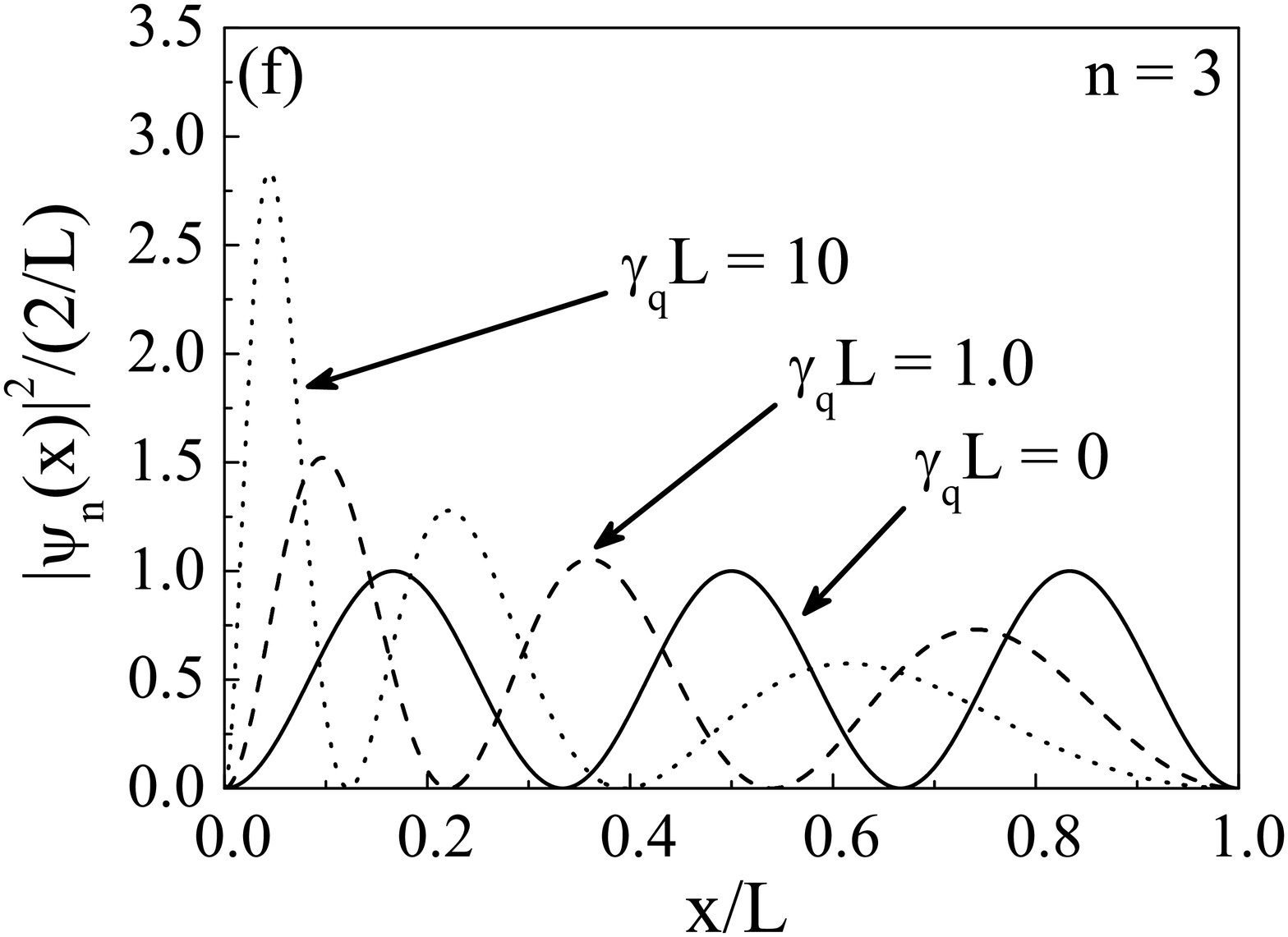}
\end{minipage}
\caption{\label{fig:psi_and_psi_quad} 
Wave functions $\psi_n (x)$ (left column)
and probability densities $|\psi_n (x)|^2 $ (right column),
conveniently scaled, for a particle confined 
in an infinite square well within a generalized space 
for different values of $\gamma_q L$ 
(indicated, usual case $\gamma_q L=0$ is shown, for comparison).
(a) and (b): $n=1$ (ground state), 
(c) and (d): $n=2$ (first exited state),
(d) and (f): $n=3$ (second exited state).
}
\end{figure}

\begin{figure}[!b]
\centering
\begin{minipage}[h]{0.48\linewidth}
\includegraphics[width=\linewidth]{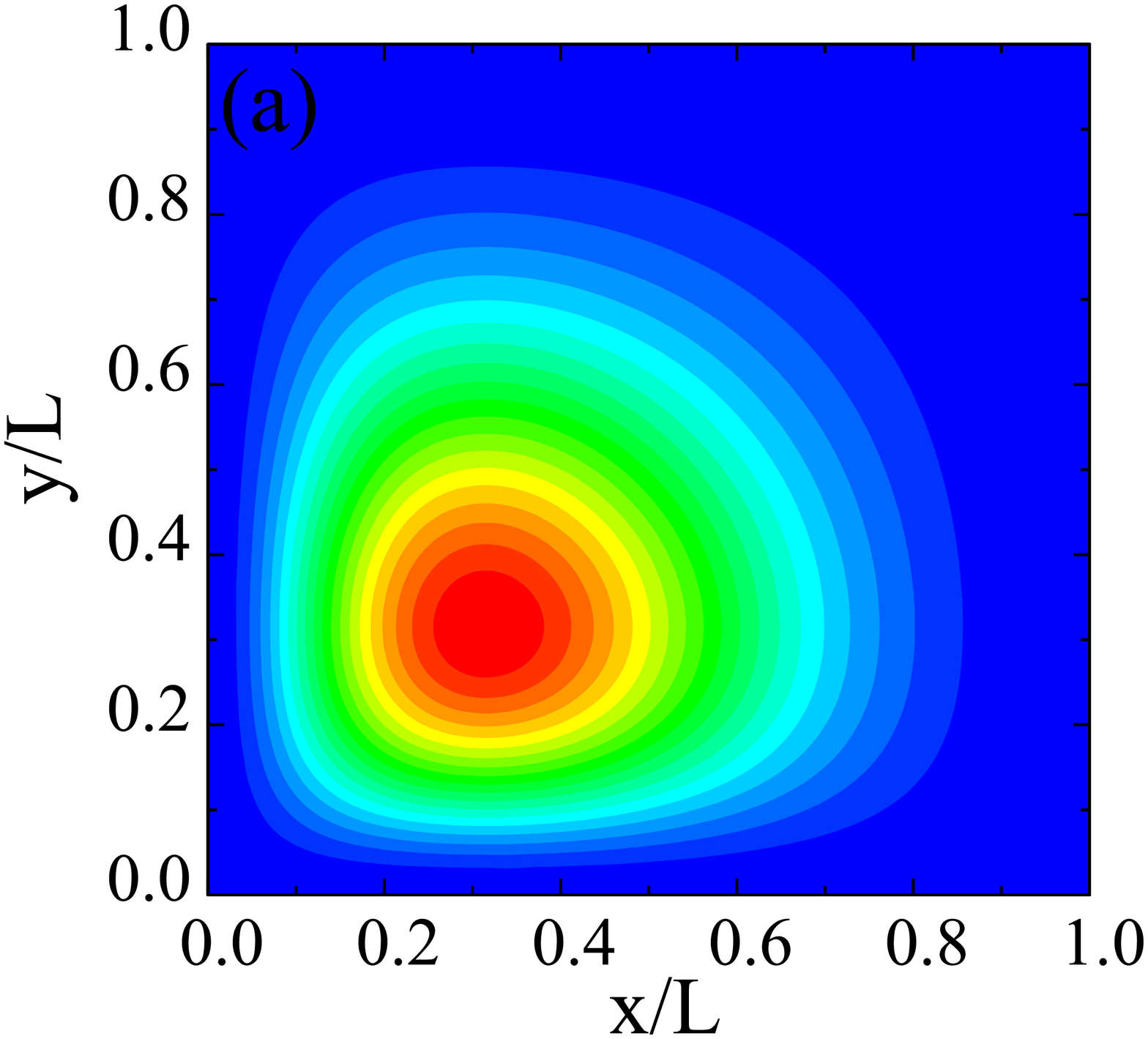}
\end{minipage}
\begin{minipage}[h]{0.48\linewidth}
\includegraphics[width=\linewidth]{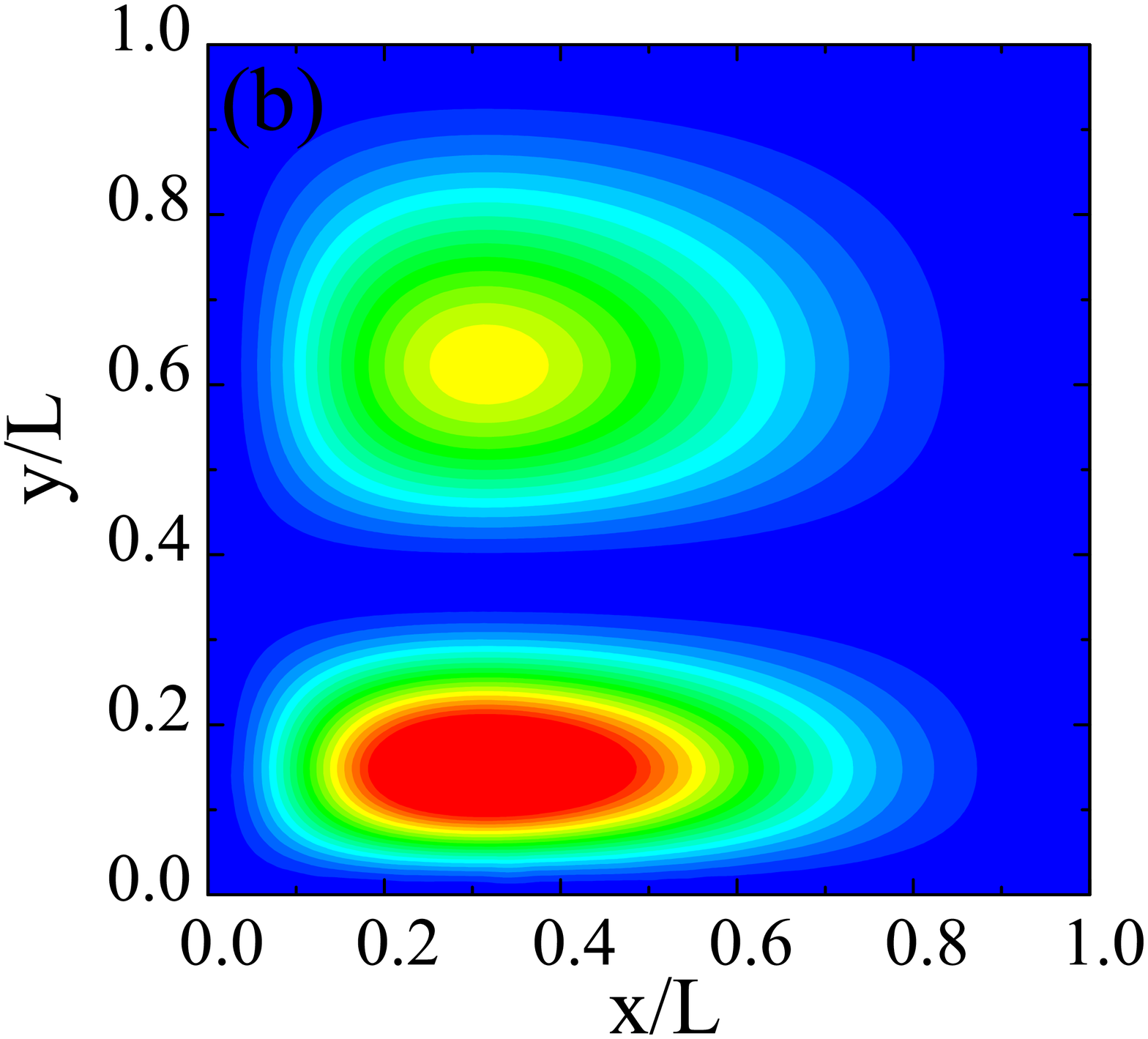}
\end{minipage}
\\
\begin{minipage}[h]{0.48\linewidth}
\includegraphics[width=\linewidth]{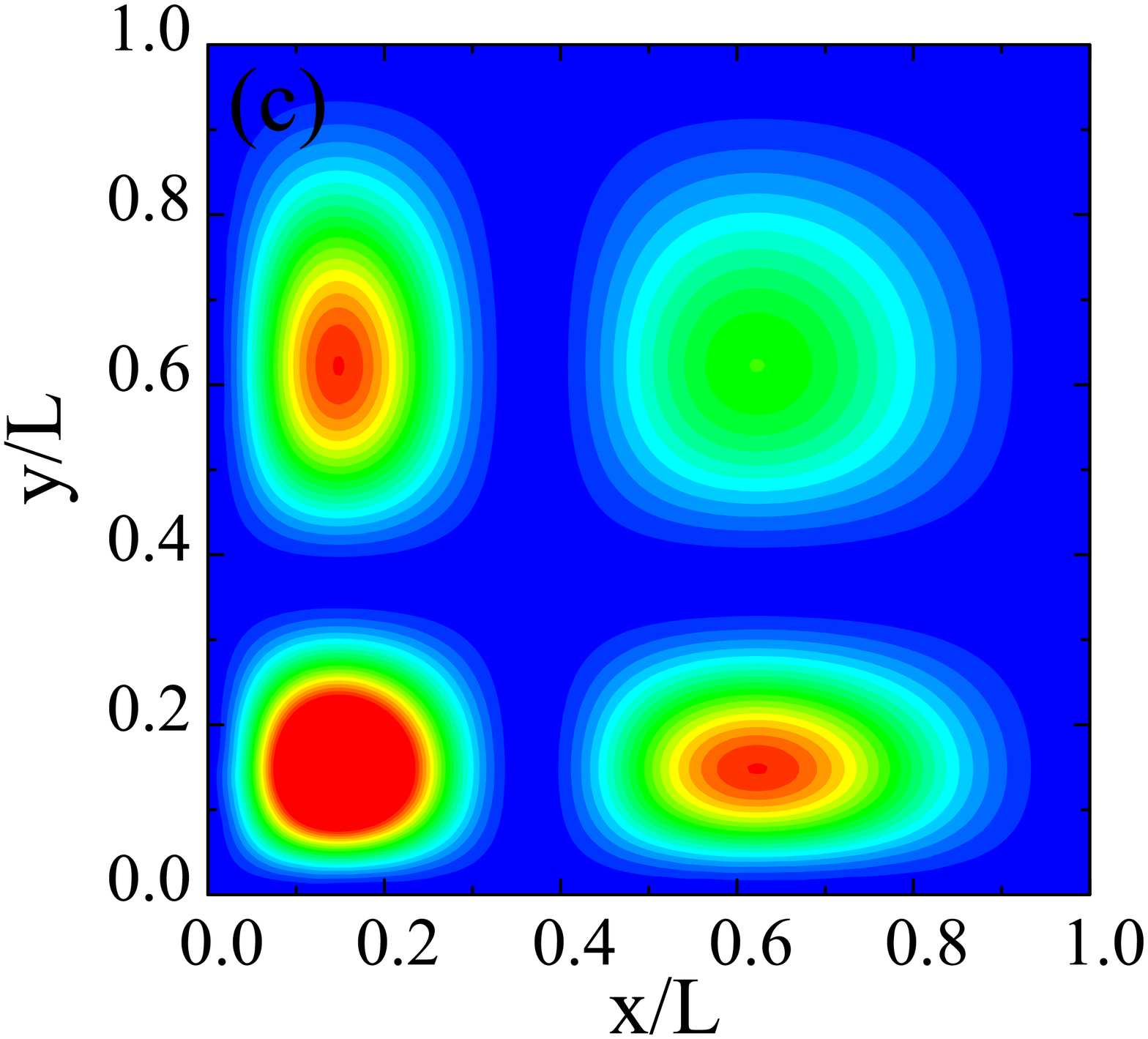}
\end{minipage}
\begin{minipage}[h]{0.48\linewidth}
\includegraphics[width=\linewidth]{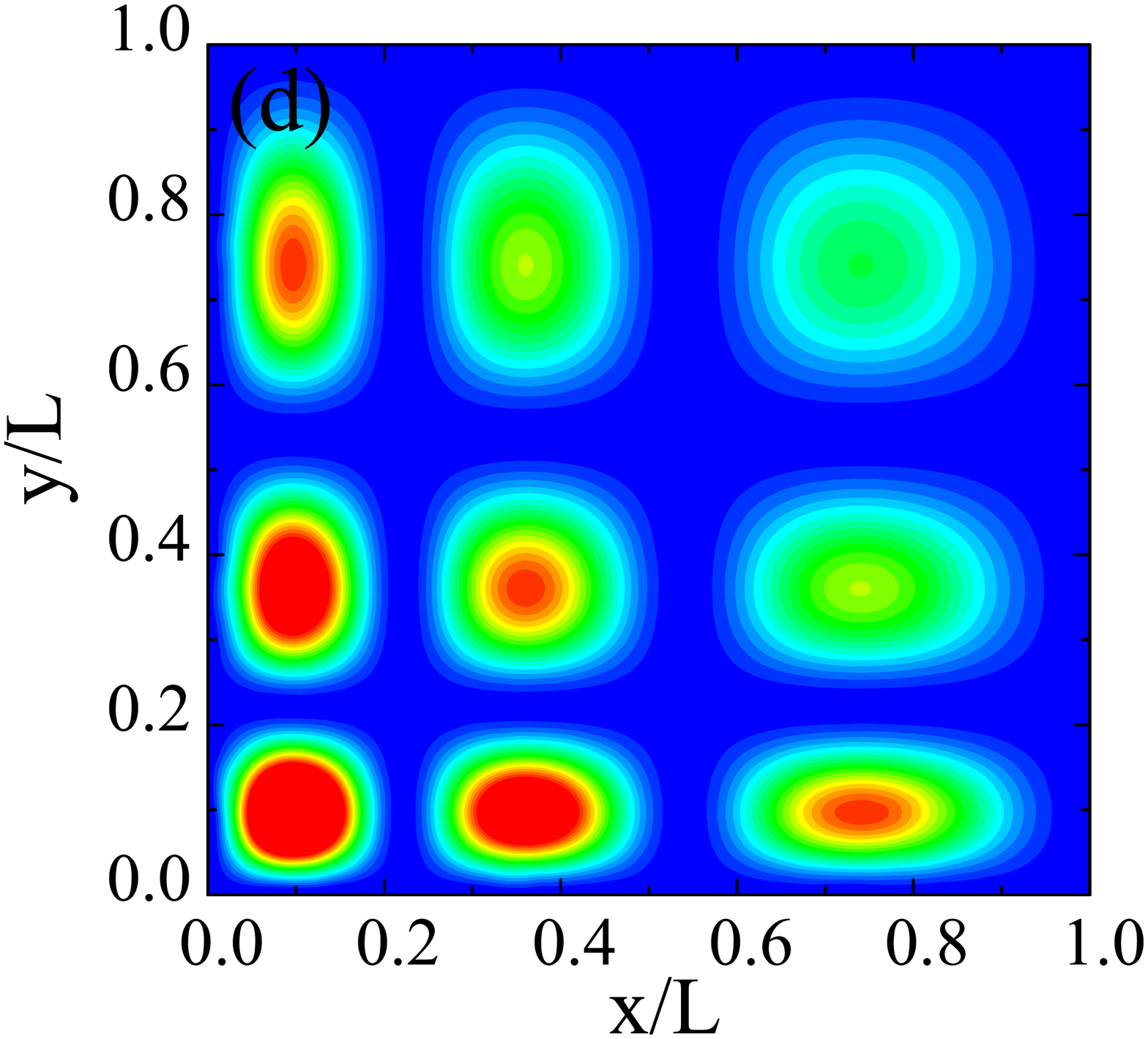}
\end{minipage}
\caption{\label{fig:psi_box} 
(Color online)
Probability density contour
$P(x,y) = |\psi_{n_1} (x) \psi_{n_2} (y)|^2$ for a particle confined
in a bidimensional box within a generalized space with $\gamma_q L = 2$, and
(a) $(n_1,n_2)=(1,1)$,
(b) $(n_1,n_2)=(1,2)$,
(c) $(n_1,n_2)=(2,2)$,
(d) $(n_1,n_2)=(3,3)$.
Color scale ranges from blue (low probabilities) to red (high probabilities).
}
\end{figure}

Figure \ref{fig:psi_and_psi_quad} shows the wave functions and their 
respective probability densities for the three states of lowest energy,
and
Figure \ref{fig:psi_box} illustrates four instances of
the probability density
$P(x,y) = |\psi_{n_1} (x) \psi_{n_2} (y)|^2$ for the case of a particle
with position-dependent mass in a bidimensional box.
It can be seen an asymmetry introduced by the position-dependent mass 
--- 
the probability to find the particle around $x=0$ increases 
as $\gamma_q L$ increases. 

\begin{figure}[!t]
\centering
\includegraphics[width=0.80\linewidth]{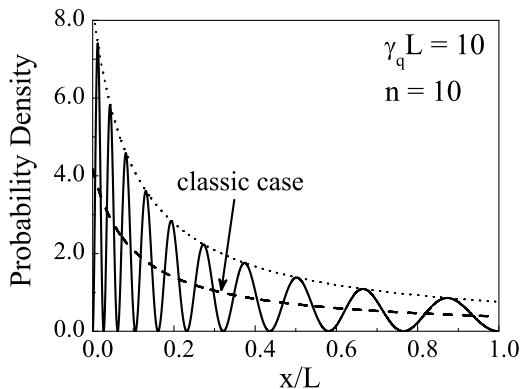}
\caption{\label{fig:psi_and_psi_quad_n10} Probability density of a particle
confined in an infinite square well of a generalized space with 
$\gamma_q L=10$ at state $n=10$.
The upper bound (dotted curve) is given by
$2\gamma_q L/[(1+\gamma_q x) \ln (1+\gamma_q L)]$.
The dashed curve is the classical case, Eq.~(\protect\ref{eq:probability_classic}).
}
\end{figure}

These results reduce to the usual problem of a particle confined in an
infinite square well in the limit $ \gamma_q \rightarrow 0 $.
We can see from Figure \ref{fig:psi_and_psi_quad_n10} that 
the average value of the quantum probability density approaches
the classical one for large quantum numbers (here exemplified with $n=10$),
consistent with the correspondence principle.

The expectation values of 
$ \langle \hat{x} \rangle $,
$ \langle \hat{x}^2 \rangle $,
$ \langle \hat{p} \rangle $, and
$ \langle \hat{p}^2 \rangle $
for the particle in a one dimensional infinite square well are given by 
\begin{subequations}
\begin{equation}
  \label{eq:x-med-quantum} 
  \langle \hat{x} \rangle = 
   \frac{\gamma_q L - \ln(1 + \gamma_q L )}{\gamma_q \ln(1 + \gamma_q L )} 
    -\frac{L \ln(1 + \gamma_q L )}{\ln^2(1 + \gamma_q L ) +(2\pi n)^2},
\end{equation}
\begin{eqnarray}
 \label{eq:x^2-med-quantum}
 \langle \hat{x}^2 \rangle &=& 
 \frac{\gamma_q^2 L^2 - 2\gamma_q L + 2\ln(1 + \gamma_q L )}{2\gamma_q^2 \ln(1 + \gamma_q L )} 
 \nonumber \\
 &&+  
 \frac{1-(1+\gamma_q L)^2 \ln(1 + \gamma_q L )}{2\gamma_q ^2 [\ln^2 (1 + \gamma_q L ) + n^2{\pi}^2]} 
 \nonumber \\
 &&+ 
 \frac{2\gamma_q L \ln(1 + \gamma_q L )}{\gamma_q ^2 [\ln^2 (1 + \gamma_q L ) + 4n^2{\pi}^2]},
\end{eqnarray}
\begin{equation}
 \label{eq:valor-expected-p}
 \langle \hat{p} \rangle = 0,
\end{equation}
\begin{eqnarray}
 \label{eq:valor-expected-p^2}
 \langle \hat{p}^2 \rangle &=&
 \hbar^2 \frac{[(1+\gamma_q L)^2 - 1]}{2(1+\gamma_q L)^2\ln (1 + \gamma_q L)}
 \frac{k_{q,n}^2}{k_{q,n}^2 + \gamma_q^2} \nonumber \\&&
 \!\!\!\!\!
 \times 
 \left[ 
       \left(k_{q,n} - \frac{i\gamma_q}{2} \right) 
       \left( k_{q,n} + \frac{i\gamma_q}{2} \right) 
       + \gamma_q^2 
 \right]\!\!.
\end{eqnarray}
\label{eq:expectation_values_quantum}
\end{subequations}

Clearly we can see that in the limit $n \rightarrow \infty $, 
Eq.'s ~(\ref{eq:expectation_values_quantum}) 
coincide with Eq's ~(\ref{eq:classic_first_second_moment}),
obtained by the analogous problem described in the classical formalism.
Also easily one can show 
that the limit $\gamma_q \rightarrow 0$
recovers the usual results 
$ \langle \hat{x} \rangle \rightarrow \frac{L}{2}  $, 
$ \langle \hat{x}^2 \rangle \rightarrow \frac{L^2}{3} - 
\frac{L^2}{2n^2\pi^2} $ and 
$ \langle \hat{p}^2 \rangle \rightarrow \hbar^2k_n^2 $
with $E_n = \hbar^2 k_n^2 /2m$ ($k_n\equiv k_{1,n}=2\pi n/L$).

\begin{figure}[!h]
\centering
\includegraphics[width=0.80\linewidth ]{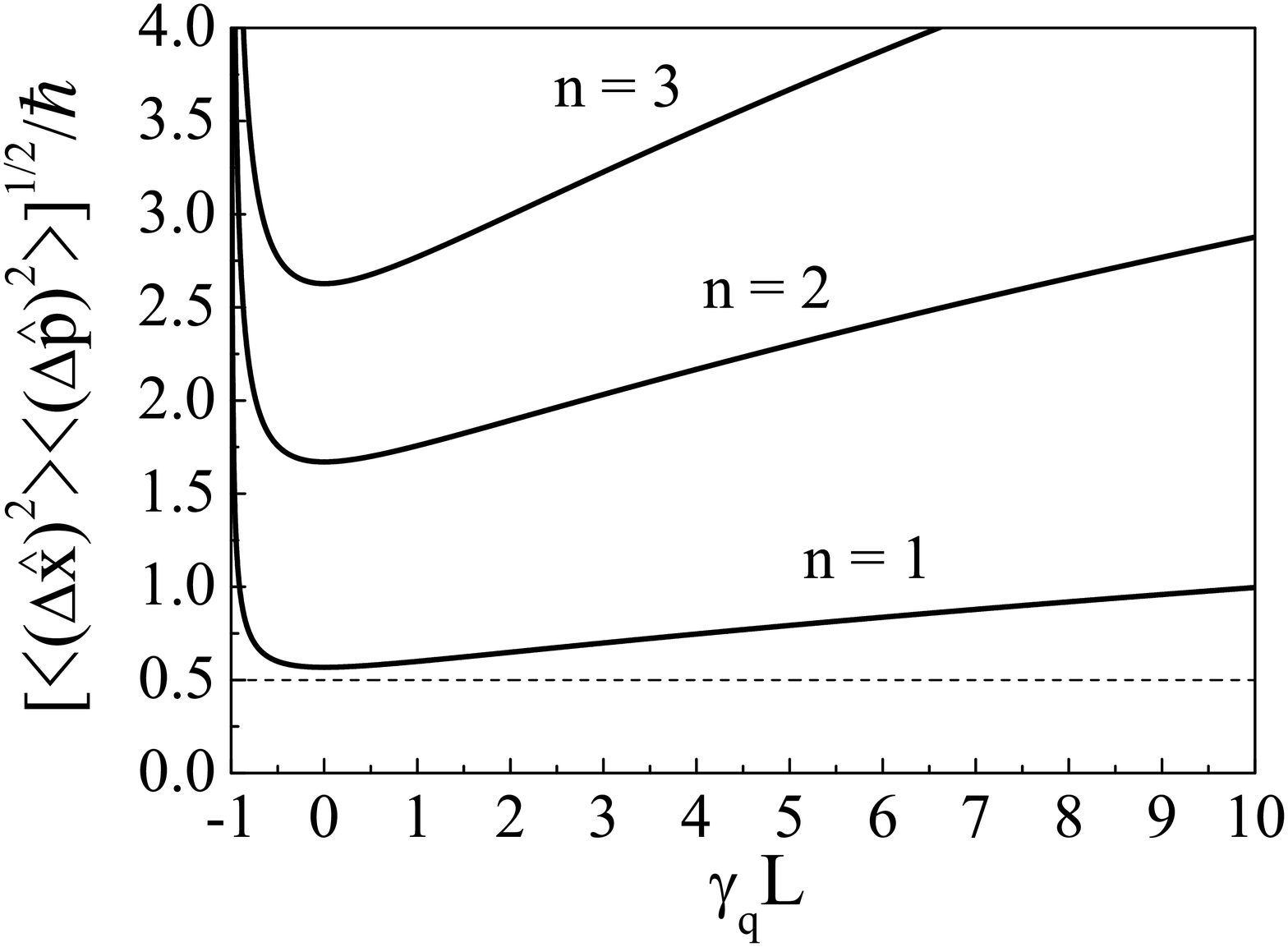}
\caption{\label{fig:incerteza} 
$[\langle (\Delta\hat{x})^2\rangle
\langle (\Delta\hat{p})^2\rangle]^{1/2}/\hbar $
for different states of a confined particle, as a function of $\gamma_q L$,
for states $n=1, 2, 3$.
}
\end{figure}

Since the operators $\hat{x}$ and $\hat{p}$ are Hermitian and 
canonically conjugated, the uncertainty relation is satisfied
for different values of $\gamma_q$, {\it i.e.}
$\langle (\Delta \hat{x})^2 \rangle \langle (\Delta \hat{p})^2 \rangle 
  \geq \hbar^2/4$
(see Figure \ref{fig:incerteza}).
Note that the product 
$\langle (\Delta \hat{x})^2 \rangle \langle (\Delta \hat{p})^2 \rangle$
is minimum for $\gamma_q = 0$.

Finally, we conclude that
the modified generalized translation operator $\hat{T}_q(\varepsilon)$ 
(Eq.\ (\ref{eq:translate-modif})) preserves the properties of 
that one introduced by \cite{costa-filho-2011},
Eq.~(\ref{eq:translate}).
The corresponding generalized linear momentum operator $\hat{p}_q$,
which is the generator of these translations,
is Hermitian, as suggested by \cite{mazharimousavi}. 
The canonical transformation 
$ (\hat{x}, \hat{p}) \rightarrow (\hat{x}_q, \hat{p}_q) $
leads  the Hamiltonian of a system with position-dependent mass 
given by $m(x) = m/(1 + \gamma_q x)^2$
to another one of a particle with constant mass.
Hermiticity permits the existence of classical analogs of the operators.
Particularly, the classical equation of motion in the phase space
may be compactly rewritten with the second dual $q$-derivative.
We have revisited the problem of a particle confined 
within an infinite square well, as discussed by 
\cite{costa-filho-2011}, \cite{mazharimousavi} and \cite{schmidt-2006}.
The results are consistent with the uncertainty and correspondence principles,
as expected, once these dynamical variables are canonical and Hermitian.

\begin{acknowledgments}
We thank C.\ Tsallis, R.\ N.\ Costa Filho and
J.\ S.\ Andrade Jr.\ for fruitful discussions.
This work was partially supported by FAPESB, through the program PRONEX
(Brazilian funding agency).
\end{acknowledgments}

\end{document}